\documentclass[12pt,preprint]{aastex}

\begin{document}

\title{Outliers in the 0Z Survey}

\author{Judith G. Cohen\altaffilmark{1},
   Norbert Christlieb\altaffilmark{2}, Ian Thompson\altaffilmark{3},
   Andrew McWilliam\altaffilmark{3} 
  \& Stephen Shectman\altaffilmark{3} }
 
\altaffiltext{1}{Palomar Observatory, California Institute of Technology, 
Pasadena, CA, USA}

\altaffiltext{2}{Zentrum fur Astronomie der Universitt Heidelberg, Germany}

\altaffiltext{3}{Observatories of the Carnegie Institution of Washington, Pasadena, CA, USA}

\begin{abstract}
We have now completed detailed abundance analyses of more than 100 stars selected
as candidate extremely metal-poor stars with [Fe/H] $< -3.0$~dex.
Of these 18 are below $-3.3$~dex on the scale of the First Stars VLT project
led by Cayrel, and 57 are below $-3.0$~dex on that scale.  Ignoring 
enhancement of carbon which ranges up to very large values, and two
C-rich stars with very high N as well, there are 0 to 3 high or low {\it{strong}}
outliers for each abundance ratio tested
from Mg to Ni.  The outliers have been checked and they are real.
Ignoring the outliers, the dispersions are in most cases approximately
consistent with the uncertainties, except those for [Sr/Fe] and [Ba/Fe],
which are much larger.  Approximately 6\% of the sample are strong
outliers in one or more elements between Mg and Ni.  This rises to
$\sim$15\% if minor outliers for these elements and strong outliers
for Sr and Ba are included.  There are 6 stars with extremely
low [Sr/Fe and [Ba/Fe], including one
which has lower [Ba/H] than Draco~119, the star found by Fulbright, Rich \& Castro
to have the lowest such ratio known previously.  There
is one extreme $r$-process star.

\end{abstract}

\section{Outliers in Abundance Ratio Trends}

Extremely metal poor (EMP) stars were presumably among the first
stars formed in the Galaxy, and hence represent in effect
a local high-redshift population.   Such stars
provide important clues to the chemical
history of our Galaxy, the role and type of early SN, the
mode of star formation in the proto-Milky Way, and the formation
of the Galactic halo.  \cite{beers05} compiled the small sample of EMP
stars known as of 2005.
The goal of our 0Z Project is to increase this sample
substantially.  

Our sample selection is based on mining the database of the Hamburg/ESO Survey
\citep{wis00} for candidate EMP stars with 
[Fe/H] $< -3.0$~dex \citep{christlieb03}.
Our abundance determination procedures are described in 
\cite{cohen04}.  The
determination of stellar parameters, measurement of equivalent widths, and detailed
abundance analyses were all carried out by J.~Cohen.

Our data in general follow the well established trends from numerous studies of Galactic
halo stars between abundance ratios [X/Fe] and overall
metallicity as measured by [Fe/H] \citep[see e.g.][]{cayrel_04,cohen04}.  
The interesting question is whether in the low metallicity
regime studied here we can detect
 the effect of only a small number of  SN contributing to a star's chemical
inventory or inhomogeneous mixing within the ISM at these early stages of formation of the Galaxy.
Thus the size of the scatter around these trends and whether
there are major outliers is of great interest.

After all the abundance analyses were completed, we 
looked for {\it{strong}} outliers, either
high or low, in
plots of [X/Fe] vs [Fe/H].  We checked these in detail.
The  Ca abundance turned out to be problematical 
in those  very C-rich stars whose spectra were obtained
prior to HIRES detector upgrade in mid-2004  and thus
included only a limited wavelength range.  In
an effort to derive Ca abundances from these early HIRES spectra, we ended up using
lines which were crowded/blended, presumably by molecular
features. This was only realized
fairly recently when we obtained additional C-star HIRES spectra extending
out to 8000~\AA\ which covered key isolated 
 Ca~I lines in the 6160~\AA\ region.  We found much lower Ca abundances from the additional
 Ca lines in these carbon stars.  Our
earlier claims in \cite{cohen06} of high Ca/Fe for some C-rich stars 
are not correct.

These abundance analyses were carried out
over a period of a decade, and some updates were made in J.~Cohen's
master list of adopted $gf$ values during that period.  The
next step, completed in Dec 2011 after the conference,
was to homogenize the $gf$ values. 

\section{Linear Fits to Abundance Ratios}

Fig. 1 shows [Ca/Fe] vs [Fe/H] for our sample.
Linear fits to the abundance ratios vs [Fe/H] where there is adequate data
for the species X were calculated, excluding
 C-rich stars and  a small number of  {\it{strong}} outliers. 
An example of these fits is shown 
for Ca in
Fig~2, where in the lower panel the included stars are shown together
with the fit (thick solid line) and the fit $\pm$0.15~dex (dashed lines). 
Note that the fit for [Ca/Fe] vs [Fe/H] is constant, [Ca/Fe] = 0.26~dex.  In the
upper panel the histogram of deviations from the linear fit is shown.
Although a number of very deviant low outliers were excluded, an 
assymetric distribution of $\delta$([Ca/Fe]) still remains, suggesting the presence
of a small tail of stars with low [Ca/Fe], though not so extreme
that the stars were rejected as strong outliers.  This is also apparent
in the lower panel, where there are four stars with [Ca/Fe] $< 0$;
these  values were not low enough for them to be rejected as strong outliers.
Current work focuses on determining whether the dispersion about these fits
is larger than the expected uncertainties.

There are six stars which are very deviant low outliers in [Ba/Fe].  One of these
has [Ba/H] below that of Draco~119, the star with the
lowest Ba abundance previously known \citep{draco119}.

\acknowledgements

We are very grateful
to the Palomar, Las Campanas, and Keck time allocation committees for
their long-term support of
this  effort.
J.~Cohen acknowledges partial support from NSF grants
AST--0507219 and AST--0908139. I.~Thompson acknowledges partial
support from NSF AST-0507325.
We are grateful to the many people  
who have worked to make the Keck Telescopes and their instruments,
and the Magellan Telescopes and their instruments,  
a reality and to operate and maintain these observatories.

{}

\clearpage

\begin{figure}
\epsscale{1.0}
\plotone{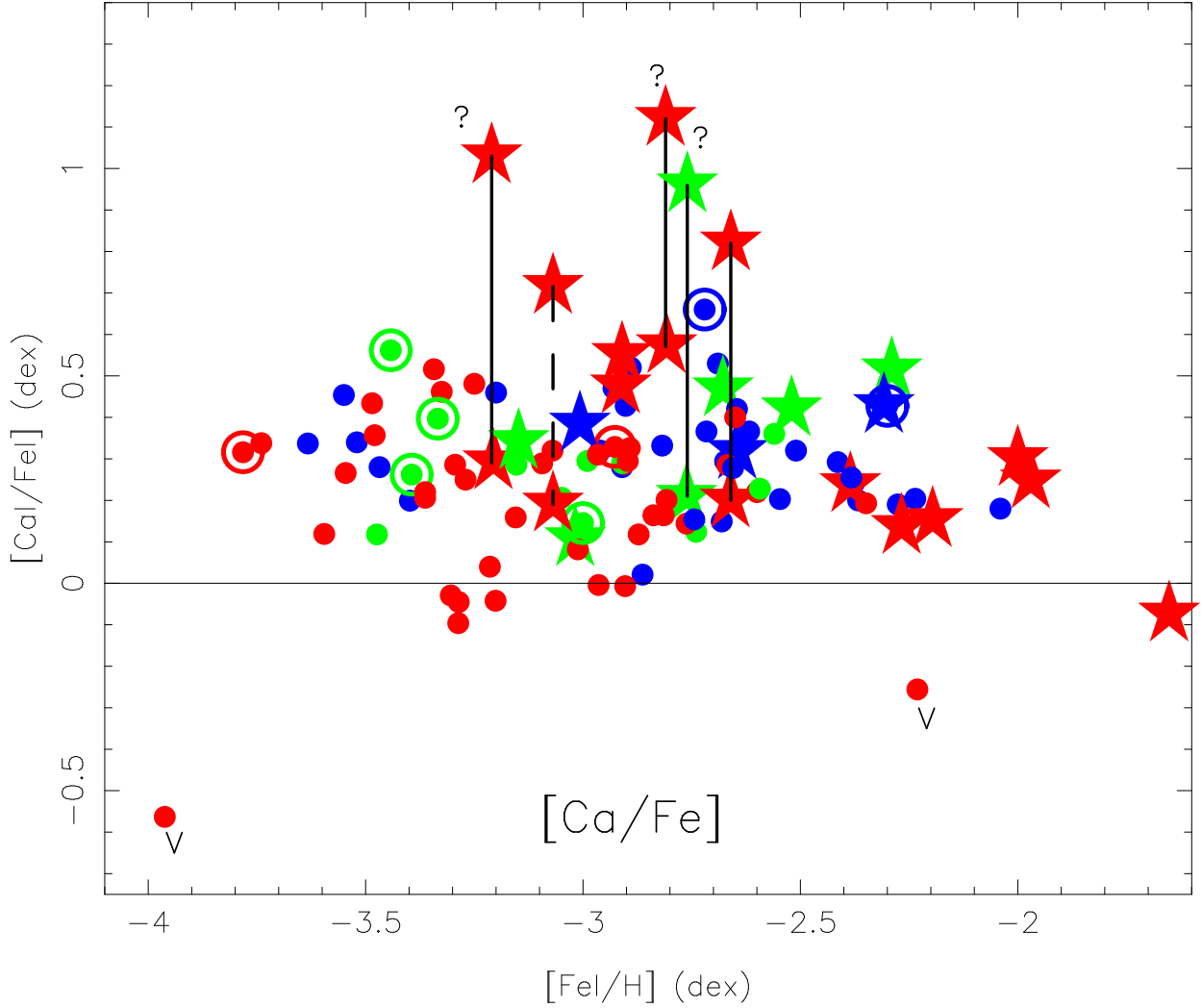}
\caption[]{[Ca/Fe] for our sample.  Stars = C-stars, circled objects
have [C/Fe] $> 1.0$~dex but no C$_2$ bands.  
Blue $6000 < T_{eff}$, green $5300 < T_{eff} < 6000$~K, red
$T_{eff} < 5300$~K. (The figures to be printed in the conference
proceedings are black-and-white.) 
Objects marked V
are outliers that have been checked.  Vertical lines 
connect the initial and final [Ca/Fe] for four C-stars which were high
outliers.  The initial high values were rated as uncertain
(marked by ?) during
the checking process.  New spectra for these four C-stars
were obtained that reach to the  6160~\AA\
region with strong Ca~I lines that is 
clear of molecular bands.  The resulting Ca abundances for
each of these C-stars  is substantially
lower; they now lie with the main distribution and are no longer
outliers.
\label{figure_cafe} }
\end{figure}

\clearpage

\begin{figure}
\epsscale{0.80}
\plotone{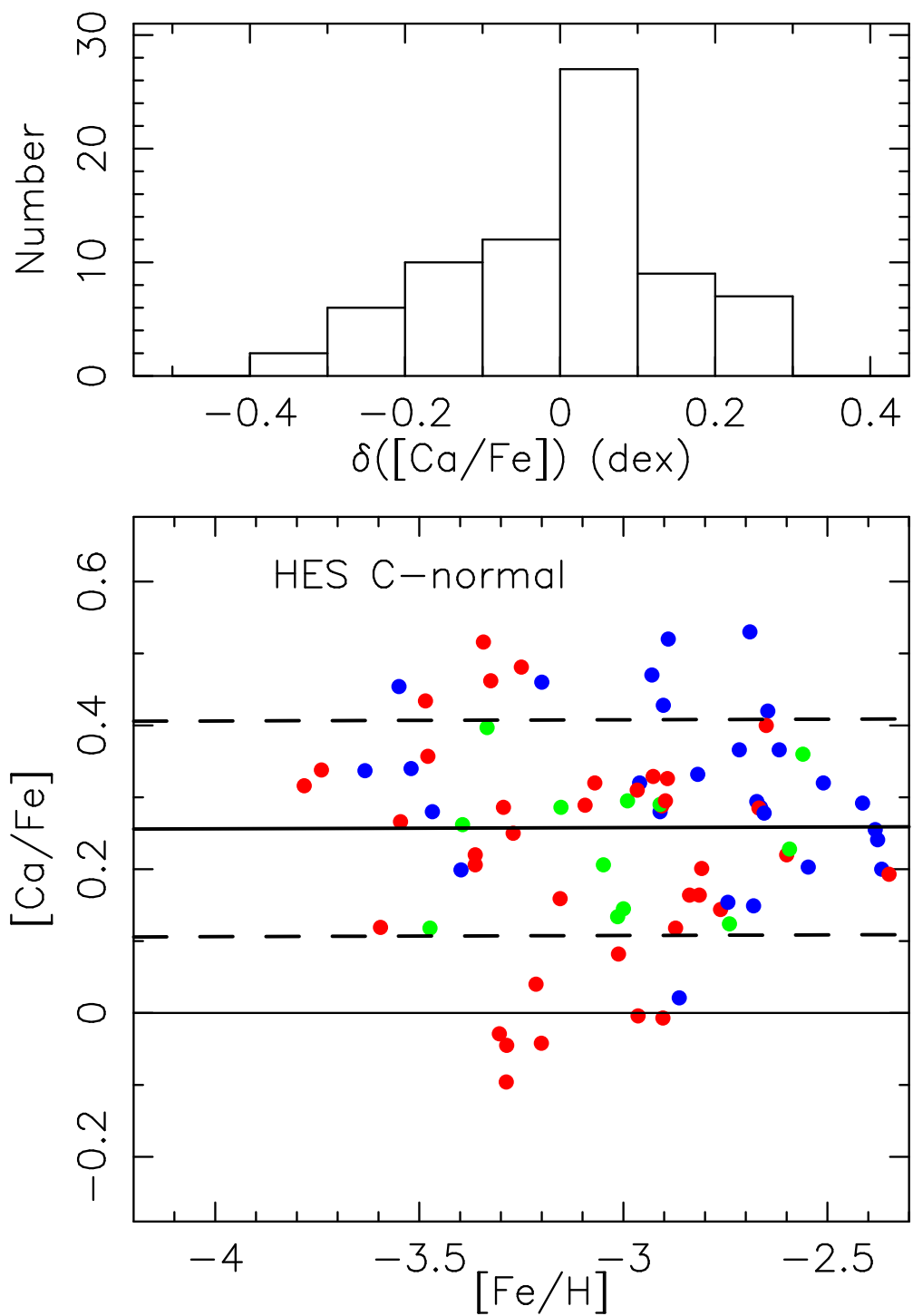}
\caption[]{Lower panel: [Ca/Fe] for stars used in the linear fit
of [Ca/Fe] with [Fe/H] for the 100 EMP candidates with high dispersion HIRES/Keck
or MIKE/Magellan specra and with detailed abundance analyses.  The symbols
are as in Fig.~1, but only the C-normal stars are shown.  The strong
outliers are also omitted.
The  linear fit is the thick solid line.  Dashed lines are the fit $\pm$0.15~dex.
Upper panel: histogram of deviations around the linear fit.  Note the 
assymetric tail towards
low [Ca/Fe] ratios.
\label{figure_cafe_fit} }
\end{figure}


\begin{thebibliography}{}

\bibitem[Beers \& Christlieb(2005)]{beers05}
Beers, T.~C. \& Christlieb, N., 2005, \araa, 43, 531

\bibitem[Cayrel et al(2004)]{cayrel_04} 
Cayrel, R. et al, 2004, \aap, 416, 1117

\bibitem[Christlieb(2003)]{christlieb03}
Christlieb, N., 2003, Rev. Mod. Astron., 16, 191

\bibitem[Cohen et al(2004)]{cohen04}
Cohen, J.~G.,et al, 2004, \apj, 612, 1107

\bibitem[Cohen et al(2006)]{cohen06}
Cohen, J.,  McWilliam, A., Shectman, S., Thompson, I., Christlieb, N.,
Ram\'irez, S., Swenson, A. \& Zickgraf, F.~J., 2006,
\aj, 132, 137


\bibitem[Fulbright, Rich \& Castro(2004)]{draco119}
Fulbright, J., Rich, R.~M. \& Castro, S., 2004, \apj, 612, 447

\bibitem[Wisotzki et al(2000)]{wis00} 
Wisotzki, L., Christlieb, N., 
Bade, N., Beckmann, V., K\"ohler, T., Vanelle, C. \& Reimers, D., 2000, 
\aap, 358, 77

\end{thebibliography}
\end{document}